\documentclass[jcp,aip,showpacs,amsmath,amssymb,unsortedaddress,12p]{revtex4}

\usepackage{epsfig}
\usepackage{graphicx}
\usepackage{chapterbib}
\usepackage{inputenc}

\begin{document}

\preprint{PREPRINT}

\title{A heuristic rule for classification of classical fluids: Master curves for Mie, Yukawa and square-well potentials}

\author{Pedro Orea}
\affiliation{Instituto Mexicano del Petr\'{o}leo, Direcci\'on de 
Investigaci\'on 
en Transformaci\'on de Hidrocarburos, Eje Central L\'{a}zaro 
C\'{a}rdenas 152, 07730 M\'{e}xico D.F., Mexico.}

\author{Szabolcs Varga}
\affiliation{Institute of Physics and Mechatronics, University of Pannonia, PO Box 158, Veszpr\'em, H-8201 Hungary.}

\author{Gerardo Odriozola}
\email{godriozo@azc.uam.mx}
\affiliation{\'Area de F\'isica de Procesos Irreversibles, Divisi\'on de 
Ciencias B\'asicas e Ingenier\'{i}a, Universidad Aut\'onoma Metropolitana, Av. 
San Pablo 180 Col. Reynosa, 02200
M\'exico D.F., Mexico}

\date{\today}

\begin{abstract}
A shift of the vapor-liquid coexistence curves by the critical value of the 
reduced second virial coefficient yields striking data collapses to define 
master curves. This is observed for the Mie, Yukawa and square-well fluids of 
different attractive ranges. This modification of the extended 
corresponding-states law of Noro and Frenkel strongly improves the outcomes from 
the van der Waals principle. Moreover, this shifted extended principle makes the 
master curves from Mie and Yukawa potentials to be one on top of the other. The 
square-well potential forms two well defined master curves, each one 
corresponding to different effective critical exponents. 
\end{abstract}


\maketitle

Simple fluids lead to approximately the same compressibility factor when 
compared at the same reduced temperature and pressure, i.~e.~all these fluids 
deviate about the same degree from the ideal behavior. In fact, the van der 
Waals equation can be written in terms of reduced variables recasting into a 
substance independent form, when the reduction of variables is made with the 
corresponding critical values. This invariance can be seen as the principle of 
corresponding states, stating that two fluids with two equal reduced variables 
must have the same other reduced properties, and so their states are said to be 
{\it corresponding}. This idea was early suggested by van der Waals and 76 years 
later was proven to be precise for spherically symmetric pairwise additive 
potentials of the form $A \varphi(r/r_o)$, where $r$ is the interparticle 
distance and $A$ and $r_o$ are constants~\cite{Pitzer39}. Hence, pairwise 
additive potentials which produce the same dimensionless $\varphi(x)$ function 
(of the dimensionless argument $x$) by rescaling distance and energy must follow 
the corresponding states principle and are said to be {\it conformal}. Nowadays, 
reducing the pair potential with $A=\epsilon$ and $r_o=\sigma$ (being $\epsilon$ 
and $\sigma$ the energy well depth and distance at which the potential crosses 
zero, respectively) is a very common and convenient practice when performing 
computer simulations due to this principle~\cite{Frenkel}.  

\begin{figure}
\resizebox{0.65\textwidth}{!}{\includegraphics{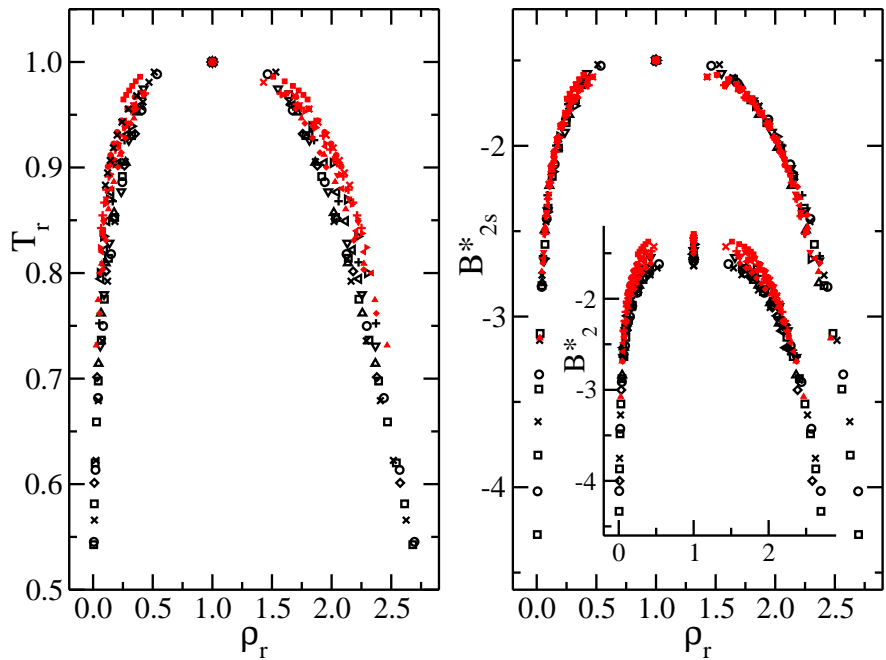}}
\caption{\label{fig1} Vapor-liquid coexistence data for the Mie and Yukawa 
potentials. Different black open symbols are employed for the Mie potential with 
different exponents $n$ and $m$. Different red filled symbols correspond to the 
attractive Yukawa with different $\kappa$. The order of the data 
as appear in the tables corresponds to squares, 
circles, diamonds, triangles-up, -left, -down, -right, plus symbols, 
and crosses in the plots. The corresponding references are also given in the 
tables. Left) $T_r-\rho_r$ chart as following the van der Waals principle. 
Right) $B_{2s}^*-\rho_r$ chart. The inset shows the $B_{2}^*-\rho_r$ chart as 
following the Noro-Frenkel proposal. The collapse of the data on the right panel 
is remarkable. }
\end{figure}

Letting aside that many body real interactions are never strictly pairwise 
additive, seldom truly spherically symmetric, and that translational and 
rotational degrees of freedom may show significant quantization effects, the 
principle is not expected to work for pair potentials of different attractive 
ranges~\cite{Guggenheim45,Leland68} (in some cases, however, it does work at the 
vicinity of the critical point for some potentials of variable 
range~\cite{Okumura00,Reyes08,Dunikov01,Weiss07,Grosfils09,Galliero09} and even 
for bulk structural and dynamical properities when the potential can be written 
as a sum of exponential functions~\cite{Dyre13,Bacher14}). This is simply 
because such potentials are not conformal. This kind of potentials is the rule 
more than the exception in the colloidal domain, since effective electrostatic 
and Hamaker interactions are omnipresent and effective polymer-mediated 
depletion interactions are frequent, contributions showing a strong range 
dependence on the properties of both phases, dispersed and 
continuous~\cite{Hunter}. With the aim of extending the applicability of the 
principle to potentials of variable attractive range, at least at the vicinity 
of the critical point, Noro and Frenkel~\cite{Noro00} suggested using the 
reduced second virial coefficient, $B_2^*$, as an additional independent 
variable. Thus, in addition to the $\sigma$ and $\epsilon$ usual quantities to 
rescale the interparticle potential, they incorporate $B_2^*$ as a measure of 
its effective attractive range. This was done in view that $B_2^*(T^*_c)$ is 
almost constant for many different shapes of $\varphi(x)$ 
functions~\cite{Vliegenthart00}(being $T^*_c$ the dimensionless critical 
temperature; we are using the following units: $T^*=k_BT/\epsilon$ for 
temperature, $\rho^*=\rho \sigma^3$ for density, $r^*=r/\sigma$ for length, and 
$u^*=u/\epsilon$ for energy). In their proposal $B_2^*(T^*_c)=-1.5$ and so, 
given a particular $\varphi(x)$, $T^*_c$ is fully determined without the need of 
a laboratory (or computational) experiment. Moreover, since master curves should 
exist for all reduced properties when represented against $B_2^*(T^*)$, one 
would obtain the desired reduced property value for any $\varphi(x)$ and $T^*$. 
That is, given $\varphi(x)$, a single variable evaluated at its critical point 
would be sufficient to predict its values at other states. This is extremely 
useful for systems having a difficult to access critical point, such as those 
governed by strong short-ranged interactions (most colloidal~\cite{Hunter} and 
globular protein suspensions~\cite{Katsonis06,Valadez12}). Nonetheless, the 
assumption $B_2^*(T^*_c)=cst.$ does not hold when considering from short to 
intermediate attractive ranges~\cite{Fu03,Rendon06,Zhou07,Largo08}. 
In fact, $B_2^*(T^*_c)$ diverges for the sticky limit of the 
attractive Yukawa potential~\cite{Gazzillo13}, which has given rise to propose 
a modification~\cite{Gazzillo13,Scholl13}. Consequently, 
this assumption leads to imperfect collapses of the reduced properties when 
represented against $B_2^*(T^*)$ (see figure 13 of reference~\cite{Valadez12} 
and the inset at the right panel of figure~\ref{fig1}). Note that the direct 
comparison with the van der Waals corresponding states framework for non 
conformal potentials is unfair, since in this last case two variables evaluated 
at the critical point are needed to make it work.  

The goal of this work is showing that, for spherically symmetric non conformal 
potentials from short to intermediate attractive range, the Noro-Frenkel 
extended framework works better than the original principle for the coexistence 
densities, when accounting for the fact that, in general, $B_2^*(T^*_c)$ is a 
function of the attractive range. For this purpose we simply shift the reduced 
density coexistence curves against $B_2^*$ the quantity $(cst.-B_2^*(T^*_c))$. 
In other words, we express the reduced densities against 
$B_{2s}^*(T^*)=B_2^*(T^*)-(B_2^*(T^*_c)-cst.)$, which forces all curves to 
coincide at the critical point as the van der Waals principle does. This way of 
comparison between the original and the extended laws turns to be fair. It also 
makes the extended framework dependent of the critical temperature, though. We 
are setting $cst.=-1.5$ to gain consistency with previous 
works~\cite{Vliegenthart00,Noro00,Katsonis06,Valadez12}, but probably $cst.=0$ 
would be a better choice.  

We are mostly taking advantage of the vast data series already published by 
several authors~\cite{Galliero09,Galicia08,delRio02,Patel05} and in previous 
works~\cite{Reyes08,Duda07,Odriozola11,Orea04}. We are, however, adding some 
extra results for the coexistence of the square-well and the attractive Yukawa 
models at very short attractive ranges, where no data are available. 
These new data are given in tables at the end of this document. For this 
purpose we have carried out replica exchange Monte Carlo 
simulations expanding the isothermal-isocoric ensemble in temperature. 
Temperatures are set below the critical point and following a geometrical 
decrease. This way acceptance rates turn similar between two adjacent replicas. 
The geometrical factor is set to obtain acceptance rates above $10\%$ in all 
cases (they are close to $20\%$). We are using rectangular parallelepiped cells 
with $L_x=L_y=8.0\sigma$ and $L_z=40.0\sigma$, while considering a liquid slab 
at least $10.0\sigma$ wide along the $z$ axis. The vapor and liquid phases have 
similar volumes and their bulk densities are not affected by the presence of the 
two interfaces. The initial configuration is set close to the vapor-liquid 
state, by randomly placing $N$ particles within the slab surrounded by vacuum. 
The center of mass is placed and kept along the simulation run at the cell 
center. Verlet list are employed to improve performance. After a sufficiently 
large equilibration time, $10^{10}$ cycles (consisting on $N$ trial 
displacements) are considered to obtain the reported densities.     

The square well potential reads
\begin{equation}
u^*(r^*)=\left \{ \begin{array}{ll} \infty, & \mbox{ for $r^* \leq 1,$ } \\
                 -1, &  \mbox{ for $1 < r^* \leq \lambda,$ } \\
                 0, & \mbox{ for $\lambda < r^* ,$ }
                \end{array} \right.
\end{equation}
where $\lambda$ is the range of the attractive well. We obtained 
new data for $\lambda=1.10$, 1.12, 1.15, 1.20, 1.25, 1.30 (table V of 
supplementary materials) and took data for $\lambda=1.50$, 1.75, 2.00, 2.30, 
2.50, 2.70, and 3.0 from elsewhere~\cite{Orea04,delRio02,Patel05} (tables VI 
and VII of supplementary materials).

The other two potentials we are considering are the Mie and Yukawa expressions. 
They are given by
\begin{equation}
u^*(r^*)= \frac{n}{(n-m)} \left(\frac{n}{m}\right)^{\frac{m}{(n-m)}}[ (1/r^*)^n-(1/r^*)^m  ]
\end{equation}
and 
\begin{equation}
u^*(r^*)=\left \{ \begin{array}{ll} \infty, & \mbox{ for $r^* \leq 1,$ } \\
                 - \frac{e^{(-\kappa (r^*-1))}}{r^*}, &  \mbox{ for $1 < r^*,$ } \\
                \end{array} \right.
\end{equation}
respectively. The softness and attractive range for the Mie expression are 
tuned by $n$ and $m$, respectively, with $n>m$. We have considered 
data for $m=6$ with $n=8$, 10, 12, 18, 20, and 32~\cite{Grosfils09,Reyes08} 
(table I at the end of this document), and also for $n=2m$ with $m=7$, 9, 
and 12~\cite{Reyes08} (table II). In the last 
expression $\kappa$ defines the attractive range of the Yukawa potential. 
We are accounting from $\kappa=3$ to 
10 with increments of unity~\cite{Galicia08,Duda07,Odriozola11}, and for 
$\kappa=15$ (tables III and IV). 

We start the discussion of the results for the Mie and Yukawa potentials. 
Results for the square-well interaction are given at the end of this work. The 
reason for doing so is that the shape of the square-well vapor-liquid 
coexistence curves is nearly cubic for $\lambda \le 1.75$ and nearly quadratic 
for $\lambda \ge 2$~\cite{Vega92,deMiguel97} when presented as a 
temperature-density chart. In other words, the effective critical 
exponent $\beta_e$ is close to $1/3$ (cubic) or $1/2$ (quadratic) depending on 
$\lambda$. This fact impedes producing a single master curve.     

Mie is also known as a generalized Lennard Jones potential since it turns into 
the Lennard Jones function when setting $n=12$ and $m=6$. Furthermore, it allows 
tuning the hardness of the core-core repulsion (by setting $n$) independently of 
the attractive tail (controlled by $m$). Thus, Mie presents no discontinuity and 
so it has not a well defined hard-core diameter. Nonetheless, it is possible to 
define an effective diameter, $\sigma_{eff}$, somewhere in between $\sigma$ and 
the distance at which the potential is at its minimum, 
$r_{min}$~\cite{Barker76}. For this potential we are taking data published by 
Galliero {\it et~al.}~\cite{Galliero09} and from previous work~\cite{Reyes08}. 
Whenever the critical density and temperature were not available we obtain them 
by considering $\beta_e=0.325$ and from the law 
of rectilinear diameters. All these data and the corresponding $B_2^*$ and 
$B_{2s}^*$ values are included in tables (see tables at the end of this 
document). As mentioned, $B_{2s}^*(T^*)$ is simply  
$B_2^*(T^*)-(B_2^*(T^*_c)-cst.)$ where 
$B_2^*(T^*)=3B_2(T^*)/(2\pi \sigma^{*3}_{eff}(T^*))$,
\begin{equation}
B_2(T^*)=2 \pi \int_0^\infty  r^{*2} [1- e^{-u^*(r^*)/T^*}] dr^*,
\end{equation}
\begin{equation}
\sigma^*_{eff}(T^*)= \int_0^\infty  [1- e^{-u'^*(r^*)/T^*}] dr^*,
\end{equation}
$u'^*(r^*)=u^*(r^*)+1$ for $r^*<r^*_{min}$ and $u'(r^*)=0$ 
otherwise~\cite{Andersen71,Barker76}. As pointed out above, $1 \leq 
\sigma^*_{eff} \leq r^*_{min}$. Note that $2\pi \sigma^{*3}_{eff}(T^*)/3$ is the 
second virial coefficient of hard spheres with $\sigma^*=\sigma^*_{eff}$.

In the left panel of figure~\ref{fig1} the coexistence is given as a 
$T_r-\rho_r$ chart as following the van der Waals framework. Here $T_r \equiv 
T^*/T^*_c$ and $\rho_r \equiv \rho^*/\rho^*_c$, being $\rho^*_c$ the critical 
density.  Mie data correspond to black open symbols. As can be seen, there is a 
relatively good collapse of the data defining a master curve. This finding has 
been already pointed out by several authors~\cite{Okumura00,Reyes08,Galliero09}. 
The data collapse is, however, improved when the coexistence is presented as a 
$B_{2s}^*-\rho_r$ chart, as following the extended principle proposal and the 
$B_{2}^*(T^*_c)$ shift. This is shown in the right panel of the same figure 
(black open symbols). The shape of this master curve is wider close to the 
critical point since at this region the relative changes of $B_{2s}^*$ are 
smaller than that of $T_r$. Note that the relationship between $B_{2s}^*$ and 
$T^*$ is not linear, thus the master curve does not fulfill the analogous of a 
rectilinear diameters law or a scaling law involving the critical exponent 
$\beta_e$. We are also including the $B_{2}^*-\rho_r$ chart as an inset in the 
right panel to highlight the strong enhance of the data collapse when accounting 
for the $B_{2}^*(T^*_c)$ shift.   

Data for the Yukawa fluid are taken from Galicia-Pimentel {\it 
et~al.}~\cite{Galicia08} and from previous works~\cite{Duda07,Odriozola11}. We 
are also including the case for $\kappa=15$ from our simulations. Contrasting 
with the Mie potential, the Yukawa expression can only tune its attractive range 
while having a discontinuity which defines its hard core. Note that in this case 
$\sigma_{eff}=\sigma=r_{min}$ irrespective of $T^*$. One would expect this 
difference to have an effect on the vapor-liquid coexistence, on the possible 
definition of a master curve, and on its possible shape. Indeed, the data 
collapse is not so good when the densities are represented in terms of the 
reduced temperature. This is observed in the left panel of figure~\ref{fig1}, 
where red solid symbols correspond to the Yukawa fluid. Furthermore, the 
vapor-liquid coexistence gets wider as decreasing the attractive range. Hence, 
the data are not scattered around a master curve but follow a defined trend. 
Finally, and as expected, the average curve (not shown) does not coincide with 
the one obtained for the Mie potential. It is wider close to the critical point 
and its extrapolation seems to cross the Mie master curve. 

A different picture arises when following the extended framework. The result is 
shown in the right panel of the same figure. Here, not only the data clearly 
collapse but also the resulting master curve coincides with that obtained for 
the Mie potential. This allows for a correspondence between different attractive 
ranges and different potentials types even for not very short attractive ranges 
(for very short attractive ranges the shape details of the pair potential are 
expected to lose relevance~\cite{Noro00,Foffi05,Foffi06} although 
this may not be the case~\cite{Gazzillo13}). Hence, it is possible 
the mapping among different potentials by matching 
$B_{2s}^*$~\cite{Noro00,Valadez12,JoseA15}, and to pick a model as a reference 
system (usually the adhesive hard-sphere potential~\cite{Baxter68}). 

We split the liquid and vapor branches of the master curve to yield an analytic 
fit. The expression for the liquid branch is given by 
$B^*_{2sl}(\rho_{rl})=-0.475(\rho_{rl}-1)^{3.1}+cst.$, which is strikingly 
simple. A mirror expression such as 
$B^*_{2sv}(\rho_{rv})=-0.563(1-\rho_{rv})^{3.1}+cst.$ can only match the vapor 
branch close to the critical point. We have found that 
$B^*_{2sv}(\rho_{rv})=-(a\rho_{rv}^3+b\rho_{rv}^{0.5})^{-1}+(a+b)^{-1}+cst.$ 
with $a=75.1$ and $b=3.71$ fits its general shape. Note that a single 
computational experiment for $T^*<T^*_c$ to obtain $\rho^*_l$ and $\rho^*_v$ is 
enough to determine the whole binodal by using these expressions. That is, by 
equating $B^*_{2sl}(\rho^*_l/\rho^*_c)$ and $B^*_{2sv}(\rho^*_v/\rho^*_c) $ one 
gets $\rho^*_c$. Then from $B^*_{2sl}(\rho^*_l/\rho^*_c)=B^*_{2s}(T^*_c)$, 
$T^*_c$ is obtained. With this knowledge (and function $u^*(r^*)$) the complete 
binodal can be built. It will be interesting to test the predictions for other 
$u^*(r^*)$ functions. 

\begin{figure}
\resizebox{0.65\textwidth}{!}{\includegraphics{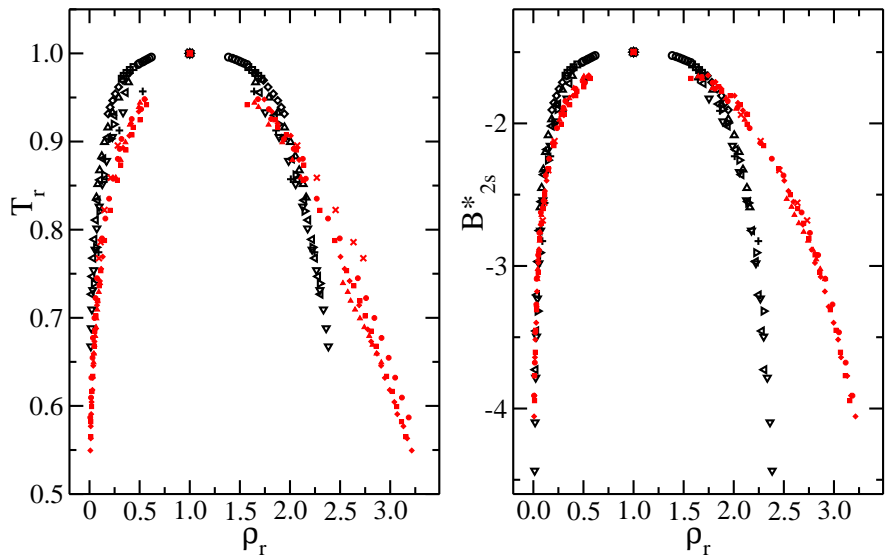}}
\caption{\label{fig2} Vapor-liquid coexistence data for the square-well 
potential. Different black open symbols are employed for different attractive 
ranges fulfilling $\lambda \le 1.75$. Red solid symbols correspond to $\lambda 
\ge 2.0$. Left) $T_r-\rho_r$ chart as by following the van der Waals principle. 
Right) $B_{2s}^*-\rho_r$ chart. Data collapse into two different master curves 
in the right panel.  }
\end{figure}

The square-well potential can be seen as a special case. Although it is 
considered the simplest model capable of producing a vapor-liquid coexistence, 
its corresponding fluid behavior is far from simple. The energy flatness of the 
well and the second discontinuity at $r^*=\lambda$ lead to a peculiar radial 
distribution function for the liquid phase~\cite{Largo05}, which in turn is 
translated to all thermodynamic bulk properties. Moreover, it shows an 
oscillatory behavior for the critical density with 
$\lambda$~\cite{Gil96,Scholl05} and a steep change in the shape of the 
coexistence curve when transiting from $\lambda=1.75$ to 
$\lambda=2.0$~\cite{Vega92,deMiguel97}. This change is probably related to the 
fact that only the first coordination shell contributes to the total potential 
energy for $\lambda \le 1.75$ and a second contributing shell appears at some 
point in between these two values~\cite{Reyes13}. Hence, the short attractive 
range suddenly turns into a not so short range, producing a jump of the 
effective critical exponent from $\beta_e \sim 0.325$ to $\beta_e \sim 0.5$. In 
other words, the nearly cubic shape of the coexistence shift to something nearly 
quadratic. This fact had raised questions on its belonging to the 3D Ising 
universality class~\cite{Vega92,deMiguel97}. In addition, $B_2^*(T^*_c)$ also 
oscillates when plotted against $\lambda$, producing values significantly 
different from $-1.5$ (see references~\cite{Fu03,Rendon06,Zhou07} and 
tables).      

Figure~\ref{fig2} shows the vapor-liquid coexistence for the square-well in 
terms of $T_r$ and $B_{2s}^*$ in the left and right panels, respectively. Here, 
we are presenting new coexistence data for very short attractive range cases 
($\lambda \leq 1.3$). Critical values well agree with those obtained from a 
mapping procedure of the grand canonical density distribution onto the universal 
Ising model distribution~\cite{Largo08}. For cases having $\lambda \geq 1.5$, 
data were taken from previous work~\cite{Orea04}, from del R\'io {\it 
et~al.}~\cite{delRio02}, and from Patel {\it et~al.}~\cite{Patel05}. We are 
employing the effective exponents $\beta_e = 0.325$ and $\beta_e = 0.5$ for 
$\lambda \leq 1.75$ and $\lambda \geq 2.0$, respectively, to obtain the critical 
parameters (all these data are given in tables). The 
short range results ($\lambda \leq 1.75$) are given as black open symbols and 
red solid symbols are used for $\lambda \geq 2.0$. Short attractive range 
results seem to collapse when plotted against $T_r$, whereas results for 
$\lambda \geq 2.0$ produce wider coexistences as increasing $\lambda$. 
Nonetheless, the collapse for $\lambda \leq 1.75$ is clearly improved when using 
$B_{2s}^*$. Moreover, the results for $\lambda \geq 2.0$ do also yield a master 
curve when plotted against $B_{2s}^*$. In this case we are not including the 
$B_2^*-\rho_r$ chart since $B_2^*(T^*_c)$ depends more strongly on the 
attractive range and so, the data collapse is poor. Unfortunately, the short 
range master curve for the square-well does not lie above the one found for Mie 
and Yukawa. The vapor branch is very similar, though (the same expression with 
$a=74.3$ and $b=3.3$ produces a good match). The liquid do not  match the simple 
shape $a(\rho_{rl}-1)^b+cst.$ 

In brief, we have compared the outcomes from the original van der Waals 
framework with those from a slight modification of the Noro-Frenkel proposal for 
non conformal potentials of variable attractive range. This has been done for 
short and intermediate attractive ranges, but for the vapor-liquid coexistence 
only. Results indicate that the extended framework works better than the 
original principle, once the corresponding shift of $B_{2}^*(T^*_c)$  from the 
proposed value is accounted for. This is pointed out, at least, for the Mie, 
Yukawa, and square-well potentials. Moreover, it is shown how the Mie and Yukawa 
potentials yield the same master curve. For the square-well case, two well 
defined master curves were obtained. It remains an open question how many master 
curves exist for simple fluids interacting for spherically symmetric pair 
potentials. In this work we have detected only three curves, but it may happen 
that new ones arise for some special shapes or ranges of the attractive 
interaction. It is also questionable whether the shifted extended principle of 
the phase coexistence can be extended for patchy colloids~\cite{Foffi07} or 
liquid crystals where the interaction is anisotropic. 

\section{Acknowledgements}

PO thank the Instituto Mexicano del Petr\'{o}leo for financial support (Project 
No D.60019). GO thanks CONACyT Project No 169125. \\



\newpage

\begingroup
\squeezetable
\begin{table}[b]
\caption{Coexistence properties and second virial coeficients for Mie fluids as 
a function of parameters $n$ and $m$. These cases correspond to $m=6$ with 
$n=8$, 10, 12, 18, 20, and 32. See the article for symbol definitions. }
\label{table1}

\begin{tabular}{ccccccc}
\hline\hline
\hspace{0.5cm}$n - m$ \hspace{0.5cm} &\hspace{0.5cm}$T^{*}$\hspace{0.5cm} & 
\hspace{0.5cm}$\rho_{L}^{*}$\hspace{0.5cm} &
\hspace{0.5cm}$\rho_{V}^{*}$\hspace{0.5cm} & \hspace{0.5cm}$B_2$\hspace{0.5cm} 
& 
\hspace{0.5cm}$B_2^{*}$\hspace{0.5cm} & \hspace{0.5cm}$B_{2s}$\hspace{0.5cm}  \\

\hline

 8-6\footnotemark[1]   & 1.000   & 0.811  & 0.0034  & -8.332  & -3.752  & 
-3.616 \\
           & 1.100   & 0.778  & 0.008  & -7.180  & -3.276  & -3.140 \\
           & 1.200   & 0.744  & 0.014  & -6.266  & -2.894  & -2.758 \\
           & 1.300  & 0.708  & 0.024  & -5.523  & -2.581  & -2.444 \\
           & 1.400  & 0.669  & 0.038  & -4.908  & -2.319  & -2.182 \\
           & 1.500   & 0.624  & 0.057  & -4.391  & -2.096  & -1.959 \\
           & 1.600   & 0.576  & 0.083  & -3.950  & -1.904  & -1.768 \\
           & 1.700   & 0.508  & 0.129  & -3.570  & -1.737  & -1.600 \\
           & 1.750  & 0.473  & 0.159  & -3.399  & -1.661  & -1.525 \\
critical point     & 1.767 & 0.309  & 0.309  & -3.343  & -1.636  & -1.500 \\ 
\hline

  10-6\footnotemark[1]  & 0.800  & 0.825  & 0.0021 & -9.307  & -4.112 & -4.024 
\\
           & 0.900  & 0.786  & 0.006 & -7.653  & -3.427 & -3.339 \\
           & 1.000  & 0.746  & 0.013 & -6.427  & -2.914 & -2.827 \\
           & 1.100  & 0.702  & 0.026 & -5.484  & -2.516 & -2.428 \\
           & 1.200  & 0.650  & 0.046 & -4.736  & -2.196 & -2.108 \\
           & 1.300  & 0.594  & 0.075 & -4.129  & -1.934 & -1.847 \\
           & 1.400  & 0.512  & 0.124 & -3.627  & -1.715 & -1.628 \\
           & 1.450  & 0.448  & 0.164 & -3.407  & -1.619 & -1.531 \\
critical point     & 1.467 & 0.306  & 0.306  & -3.336  & -1.588 & -1.500 \\ 
\hline

  12-6\footnotemark[1]  & 0.700  & 0.842  & 0.0015 & -9.865  & -4.334 & -4.277 
\\
           & 0.750  & 0.820  & 0.003 & -8.746  & -3.869 & -3.812 \\
           & 0.800  & 0.798  & 0.006 & -7.821  & -3.482 & -3.425 \\
           & 0.850  & 0.775  & 0.009 & -7.044  & -3.156 & -3.099 \\
           & 0.900  & 0.752  & 0.014 & -6.382  & -2.876 & -2.819 \\
           & 0.950  & 0.726  & 0.022 & -5.812  & -2.634 & -2.577 \\
           & 1.000  & 0.699  & 0.030 & -5.316  & -2.423 & -2.366 \\
           & 1.050  & 0.672  & 0.041 & -4.880  & -2.236 & -2.179 \\
           & 1.100  & 0.638  & 0.057 & -4.495  & -2.070 & -2.013 \\
           & 1.150  & 0.602  & 0.077 & -4.152  & -1.922 & -1.865 \\
           & 1.200  & 0.560  & 0.101 & -3.845  & -1.788 & -1.731 \\
           & 1.250  & 0.515  & 0.129 & -3.568  & -1.667  & -1.610 \\
critical point     & 1.300  & 0.312  & 0.312  & -3.318  & -1.557  & -1.500 \\
\hline

 18-6\footnotemark[2]  & 0.750  & 0.781  & 0.015 & -6.310  & -2.839  & -2.803 \\
           & 0.800  & 0.745  & 0.022 & -5.595  & -2.530  & -2.494 \\
           & 0.850  & 0.705  & 0.035 & -4.995  & -2.269  & -2.234 \\
           & 0.900  & 0.666  & 0.058 & -4.483  & -2.046  & -2.011 \\
           & 0.950  & 0.615  & 0.090 & -4.042  & -1.853  & -1.818 \\
           & 1.000  & 0.565  & 0.122 & -3.658  & -1.684  & -1.649 \\
critical point     & 1.050  & 0.330   & 0.330  & -3.321  & -1.535  & -1.500 \\ 
\hline

 20-6\footnotemark[1]  & 0.750  & 0.772 & 0.020 & -5.852 & -2.643  & -2.674 \\
           & 0.800  & 0.735 & 0.030 & -5.177 & -2.349  & -2.380 \\
           & 0.850  & 0.695 & 0.045 & -4.610 & -2.101  & -2.132 \\
           & 0.900  & 0.643 & 0.077 & -4.127 & -1.889  & -1.920 \\
           & 0.950  & 0.591 & 0.098 & -3.710 & -1.706  & -1.736 \\
           & 1.000  & 0.505 & 0.138 & -3.348 & -1.545  & -1.576 \\
critical point     & 1.026 & 0.326  & 0.326  & -3.177 & -1.469  & -1.500 \\ 
\hline

 32-6\footnotemark[2]  & 0.650  & 0.809  & 0.018 & -5.798  & -2.638  & -2.646 \\
           & 0.700  & 0.759  & 0.031 & -4.996  & -2.282  & -2.290 \\
           & 0.750  & 0.702  & 0.056 & -4.343  & -1.991  & -1.999 \\
           & 0.800  & 0.632  & 0.090 & -3.802  & -1.749  & -1.757 \\
           & 0.825 & 0.581  & 0.124 & -3.564  & -1.642  & -1.650 \\
critical point     & 0.864 & 0.340   & 0.340  & -3.231  & -1.492  & -1.500 \\ 
\hline

\end{tabular}
\footnotetext[1]{Data taken from J. Chem. Phys. {\bf 130}, 104704 (2009)}
\footnotetext[2]{Data taken from Phys. Lett. A {\bf 372}, 7024 (2008)}
\end{table}
\endgroup

\begingroup
\squeezetable
\begin{table}
\caption{Coexistence properties and second virial coeficients for Mie fluids as 
a function of parameters $n$ and $m$ (continued from TABLE \ref{table1}). These 
cases correspond to $n=2m$ with $m=7$, 9, and 12. See the article for 
symbol definitions.}
\label{table2}

\begin{tabular}{ccccccc}
\hline\hline
\hspace{0.5cm}$n - m$ \hspace{0.5cm} &\hspace{0.5cm}$T^{*}$\hspace{0.5cm} & 
\hspace{0.5cm}$\rho_{L}^{*}$\hspace{0.5cm} &
\hspace{0.5cm}$\rho_{V}^{*}$\hspace{0.5cm} & \hspace{0.5cm}$B_2$\hspace{0.5cm} 
& 
\hspace{0.5cm}$B_2^{*}$\hspace{0.5cm} & \hspace{0.5cm}$B_{2s}$\hspace{0.5cm}  \\

 14-7\footnotemark[2]  & 0.600  & 0.842  & 0.0035 & -9.109  & -4.000  & -3.952 
\\
           & 0.700  & 0.778  & 0.011  & -6.751  & -3.002  & -2.954 \\
           & 0.800  & 0.713  & 0.035  & -5.215  & -2.346  & -2.298 \\
           & 0.850  & 0.668  & 0.056  & -4.634  & -2.096  & -2.047 \\
           & 0.900 & 0.615  & 0.086  & -4.140  & -1.882  & -1.834 \\
           & 0.930  & 0.575  & 0.115  & -3.878  & -1.768  & -1.720 \\
critical point     & 0.996  & 0.330   & 0.330   & -3.360  & -1.548  & -1.500 \\ 
\hline

 18-9\footnotemark[2] & 0.580  & 0.802  & 0.024 & -5.683  & 2.537  & -2.527  \\
           & 0.600  & 0.783  & 0.035 & -5.260  & 2.353  & -2.343  \\
           & 0.620  & 0.762  & 0.042 & -4.879  & 2.188  & -2.178  \\
           & 0.640  & 0.736  & 0.054 & -4.536  & 2.038  & -2.028  \\
           & 0.660  & 0.698  & 0.076 & -4.225  & 1.902  & -1.892  \\
           & 0.680  & 0.655  & 0.104 & -3.941  & 1.778  & -1.768 \\
           & 0.700  & 0.601  & 0.145 & -3.682  & 1.664  & -1.654  \\
critical point     & 0.730  & 0.360   & 0.360 & -3.332  & 1.510  & -1.500  \\ 
\hline

24-12\footnotemark[2]  & 0.460  & 0.897  & 0.023 & -5.706  & -2.564 & -2.662  \\
           & 0.480  & 0.865  & 0.034 & -5.086  & -2.290 & -2.388 \\
           & 0.500  & 0.831  & 0.053 & -4.553  & -2.054 & -2.152 \\
           & 0.520 & 0.785  & 0.075 & -4.091  & -1.848 & -1.946 \\
           & 0.540  & 0.715  & 0.125 & -3.686  & -1.668 & -1.766 \\
 critical point     & 0.575 & 0.390  & 0.390  & -3.089  & -1.402 & -1.500 \\ 
\hline

\end{tabular}
\footnotetext[2]{Data taken from Phys. Lett. A {\bf 372}, 7024 (2008)}
\end{table}
\endgroup

\begingroup
\squeezetable
\begin{table}
\caption{Coexistence properties and second virial coeficients for atractive 
Yukawa fluids with different $\kappa$. See the article for symbol definitions.}
\label{table3}

\begin{tabular}{ccccccc}
\hline\hline
\hspace{0.5cm}$\kappa$ \hspace{0.5cm} &\hspace{0.5cm}$T^{*}$\hspace{0.5cm} & 
\hspace{0.5cm}$\rho_{L}^{*}$\hspace{0.5cm} &
\hspace{0.5cm}$\rho_{V}^{*}$\hspace{0.5cm} & \hspace{0.5cm}$B_2$\hspace{0.5cm} 
& 
\hspace{0.5cm}$B_2^{*}$\hspace{0.5cm} & \hspace{0.5cm}$B_{2s}$\hspace{0.5cm}  \\

\hline
 
 3.0\footnotemark[3] & 0.550   &  0.847  &  0.0187  & -5.631  & -2.688  & 
-2.686 
 \\
        & 0.580   &  0.809  &  0.029  & -5.042  & -2.407  & -2.405  \\
        & 0.600   &  0.780  &  0.038  & -4.697  & -2.242  & -2.240  \\
        & 0.650   &  0.698  &  0.074  & -3.963  & -1.892  & -1.890  \\ 
        & 0.660   &  0.679  &  0.085  & -3.835  & -1.830  & -1.829  \\ 
        & 0.670   &  0.656  &  0.097  & -3.711  & -1.772  & -1.770  \\
        & 0.680   &  0.633  &  0.112  & -3.594  & -1.716  & -1.714  \\
        & 0.690   &  0.607  &  0.129  & -3.480  & -1.661  & -1.660  \\ 
        & 0.700   &  0.576  &  0.152  & -3.371  & -1.609  & -1.608  \\ 
critical point      & 0.722  &  0.357  &  0.357   & -3.146  & -1.502  & -1.500  
 
 \\  \hline

 4.0\footnotemark[4] & 0.425 & 0.937  &  0.010 & -6.440     & -3.075      & 
-3.126  \\
        & 0.450  & 0.898  &  0.016 & -5.628     & -2.687      & -2.738  \\
        & 0.470  & 0.863  &  0.027 & -5.078     & -2.425      & -2.476  \\
        & 0.485 & 0.838  &  0.035 & -4.714     & -2.251      & -2.302  \\
        & 0.500  & 0.805  &  0.049 & -4.384     & -2.093      & -2.144  \\
        & 0.515 & 0.771  &  0.066 & -4.084     & -1.950      & -2.001  \\
        & 0.530  & 0.736  &  0.086 & -3.810     & -1.819      & -1.870  \\
        & 0.550  & 0.664  &  0.130 & -3.480     & -1.661      & -1.712  \\
critical point      & 0.581 & 0.380  & 0.380 & -3.034     & -1.448      & 
-1.500 
 \\ \hline
         
 5.0\footnotemark[4] & 0.400  &  0.932     &   0.0227 &  -5.403   & -2.580      
& -2.659  \\
        & 0.410  &  0.906  &   0.029 &  -5.065   & -2.418      & -2.498  \\
        & 0.420  &  0.880  &   0.036 &  -4.755   & -2.270      & -2.350  \\
        & 0.440  &  0.832  &   0.059 &  -4.207   & -2.009      & -2.087  \\
        & 0.460  &  0.776  &   0.087 &  -3.737   & -1.784      & -1.863  \\
        & 0.470  &  0.734  &   0.111 &  -3.527   & -1.684      & -1.763  \\
        & 0.480  &  0.680  &   0.140 &  -3.330   & -1.590      & -1.669  \\
        & 0.485 &  0.640   &   0.170 &  -3.237   & -1.546      & -1.625  \\
critical point      & 0.500 &  0.399   &   0.399  &  -2.976   & -1.421      & 
-1.500  \\ \hline

 6.0\footnotemark[4] & 0.380  &  0.925   & 0.037   & -4.730   & -2.258     & 
-2.361  \\
        & 0.390   &  0.892   & 0.046    & -4.392   & -2.097     & -2.199  \\
        & 0.400   &  0.852   & 0.060    & -4.084   & -1.950     & -2.053  \\
        & 0.420   &  0.770   & 0.110    & -3.545   & -1.693     & -1.795  \\ 
        & 0.430   &  0.720   & 0.140    & -3.308   & -1.579     & -1.682  \\ 
        & 0.435  &  0.680   & 0.170    & -3.197   & -1.526     & -1.629  \\
critical point     & 0.448  & 0.416    & 0.416     & -2.927   & -1.398     & 
-1.500  \\  \hline

 7.0\footnotemark[5] & 0.340   &  0.988   &  0.025   & -5.202   & -2.483     & 
-2.622  \\
        & 0.346   &  0.972   &  0.030   & -4.952   & -2.365     & -2.503  \\
        & 0.351   &  0.955   &  0.036   & -4.715   & -2.251     & -2.390  \\
        & 0.357   &  0.935   &  0.043   & -4.488   & -2.143     & -2.282  \\ 
        & 0.363   &  0.917   &  0.052   & -4.272   & -2.040     & -2.178  \\ 
        & 0.369   &  0.893   &  0.063   & -4.066   & -1.941     & -2.080  \\
        & 0.375   &  0.870   &  0.076   & -3.868   & -1.847     & -1.986  \\
        & 0.381   &  0.839   &  0.092   & -3.679   & -1.756     & -1.895  \\ 
        & 0.387   &  0.806   &  0.113   & -3.498   & -1.670     & -1.809  \\
        & 0.394   &  0.764   &  0.142   & -3.325   & -1.588     & -1.726  \\ 
        & 0.400   &  0.685   &  0.183   & -3.159   & -1.508     & -1.647  \\ 
critical point     & 0.413   &  0.431   &  0.431   & -2-850   & -1.361     & 
-1.500    \\  \hline

\end{tabular}
\footnotetext[3]{Data taken from Fluid Phase Equilib. {\bf 265}, 205 (2008)}
\footnotetext[4]{Data taken from J. Chem. Phys. {\bf 126}, 224510 (2007)}
\footnotetext[5]{Data taken from J. Chem. Phys. {\bf 134}, 154702 (2011)}
\end{table}
\endgroup

\begingroup
\squeezetable
\begin{table}
\caption{Coexistence properties and second virial coeficients for atractive 
Yukawa fluids with different $\kappa$ (continued from TABLE \ref{table3}). See 
the article for symbol definitions.}
\label{table4}

\begin{tabular}{ccccccc}
\hline\hline
\hspace{0.5cm}$\kappa$ \hspace{0.5cm} &\hspace{0.5cm}$T^{*}$\hspace{0.5cm} & 
\hspace{0.5cm}$\rho_{L}^{*}$\hspace{0.5cm} &
\hspace{0.5cm}$\rho_{V}^{*}$\hspace{0.5cm} & \hspace{0.5cm}$B_2$\hspace{0.5cm} 
& 
\hspace{0.5cm}$B_2^{*}$\hspace{0.5cm} & \hspace{0.5cm}$B_{2s}$\hspace{0.5cm}  \\

\hline
  
 8.0\footnotemark[5] & 0.325   & 0.997    & 0.030    & -4.919   & -2.349     & 
-2.516  \\
        & 0.330   & 0.980    & 0.036    & -4.700   & -2.244     & -2.411  \\
        & 0.335   & 0.966    & 0.042    & -4.491   & -2.144     & -2.311  \\
        & 0.339   & 0.948    & 0.049    & -4.290   & -2.048     & -2.215  \\ 
        & 0.344   & 0.929    & 0.058    & -4.098   & -1.956     & -2.123  \\ 
        & 0.349   & 0.906    & 0.069    & -3.913   & -1.868     & -2.035  \\
        & 0.354   & 0.881    & 0.083    & -3.736   & -1.783     & -1.950  \\
        & 0.359   & 0.855    & 0.100    & -3.566   & -1.702     & -1.869  \\ 
        & 0.364   & 0.822    & 0.121    & -3.403   & -1.624     & -1.792  \\ 
        & 0.369   & 0.772    & 0.150    & -3.246   & -1.550     & -1.717  \\
        & 0.375   & 0.696    & 0.192    & -3.095   & -1.478     & -1.645  \\
critical point     & 0.387  & 0.443    & 0.443   & -2.791   & -1.332     & 
-1.500   \\  \hline

 9.0\footnotemark[5] & 0.321   & 0.976    & 0.047    & -4.308   & -2.057     & 
-2.225  \\
        & 0.325   & 0.957    & 0.055    & -4.120   & -1.967     & -2.135  \\
        & 0.330   & 0.937    & 0.065    & -3.939   & -1.880     & -2.049  \\
        & 0.334   & 0.914    & 0.078    & -3.765   & -1.798     & -1.966  \\ 
        & 0.338   & 0.887    & 0.092    & -3.598   & -1.718     & -1.886  \\ 
        & 0.343   & 0.856    & 0.110    & -3.437   & -1.641     & -1.809  \\
        & 0.347   & 0.822    & 0.134    & -3.282   & -1.567     & -1.736  \\
        & 0.352   & 0.767    & 0.166    & -3.134   & -1.496     & -1.665  \\ 
        & 0.357   & 0.648    & 0.213    & -2.991   & -1.428     & -1.596  \\ 
critical point     & 0.364   & 0.454    & 0.454    & -2.789   & -1.331     & 
-1.500    \\  \hline

 10.0\footnotemark[5] & 0.313   &  0.967   & 0.058    & -4.024   & -1.921     & 
-2.113  \\
        & 0.316   &  0.948   & 0.068    & -3.857   & -1.841     & -2.033  \\
        & 0.320   &  0.928   & 0.080    & -3.696   & -1.765     & -1.956  \\
        & 0.324   &  0.902   & 0.094    & -3.541   & -1.691     & -1.882  \\ 
        & 0.328   &  0.876   & 0.112    & -3.392   & -1.619     & -1.811  \\ 
        & 0.332   &  0.840   & 0.135    & -3.249   & -1.551     & -1.743  \\
        & 0.336   &  0.790   & 0.163    & -3.110   & -1.485     & -1.677  \\
critical point     & 0.348  & 0.465   & 0.465    & -2.740   & -1.308     & 
-1.500    \\  \hline

 15.0\footnotemark[6] & 0.2822   &  0.902   & 0.128    & -3.167   & -1.512     
& 
-1.724  \\
        & 0.2834   &  0.885   & 0.139   & -3.106   & -1.483     & -1.695  \\ 
        & 0.2847   &  0.865   & 0.152   & -3.046   & -1.454     & -1.666  \\
        & 0.2859   &  0.841   & 0.165   & -2.988   & -1.426     & -1.638  \\
        & 0.2872   &  0.806   & 0.179   & -2.930   & -1.399     & -1.611  \\ 
        & 0.2885   &  0.758   & 0.198   & -2.873   & -1.372     & -1.584  \\ 
critical point     & 0.2926   &  0.502   & 0.502   & -2.697   & -1.288     & 
-1.500  \\ \hline

\end{tabular}
\footnotetext[5]{Data taken from J. Chem. Phys. {\bf 134}, 154702 (2011)}
\footnotetext[6]{This work}
\end{table}
\endgroup

\begingroup
\squeezetable
\begin{table}
\caption{Coexistence properties and second virial coeficients for square-well 
fluids with different $\lambda$. See the article for symbol definitions.}

\label{table5}

\begin{tabular}{ccccccc}
\hline\hline
\hspace{0.5cm}$\lambda$ \hspace{0.5cm} &\hspace{0.5cm}$T^{*}$\hspace{0.5cm} & 
\hspace{0.5cm}$\rho_{L}^{*}$\hspace{0.5cm} &
\hspace{0.5cm}$\rho_{V}^{*}$\hspace{0.5cm} & \hspace{0.5cm}$B_2$\hspace{0.5cm} 
& 
\hspace{0.5cm}$B_2^{*}$\hspace{0.5cm} & \hspace{0.5cm}$B_{2s}$\hspace{0.5cm}  \\

\hline
    
  1.10\footnotemark[6] 
         & 0.4710  & 0.747 & 0.230 & -3.006  & -1.435 & -1.573 \\
         & 0.4717  & 0.733 & 0.238 & -2.988  & -1.427 & -1.564 \\
         & 0.4726  & 0.719 & 0.252 & -2.971  & -1.419 & -1.556 \\
         & 0-4730  & 0.707 & 0.261 & -2.954  & -1.410 & -1.548 \\
         & 0.4737  & 0.693 & 0.272 & -2.937  & -1.402 & -1.540 \\
         & 0.4743  & 0.677 & 0.283 & -2.920  & -1.394 & -1.532 \\
         & 0.4750  & 0.659 & 0.293 & -2.903  & -1.386 & -1.524 \\
critical point     & 0.4770  & 0.476 & 0.476  & -2.853  & -1.362 & -1.500 \\ 
\hline

  1.12\footnotemark[6] 
         & 0.5021  & 0.806 & 0.151 & -3.272  & -1.562 & -1.665 \\
         & 0.5038  & 0.791 & 0.160 & -3.229  & -1.542 & -1.645 \\
         & 0.5055  & 0.777 & 0.173 & -3.188  & -1.522 & -1.626 \\
         & 0.5073  & 0.762 & 0.189 & -3.147  & -1.503 & -1.606 \\
         & 0.5090  & 0.745 & 0.206 & -3.106  & -1.483 & -1.586 \\
critical point     & 0.5170  & 0.470  & 0.470  & -2.925  & -1.397 & -1.500 \\   
\hline

  1.15\footnotemark[6] 
         & 0.536  & 0.863 & 0.082 & -3.873  & -1.849 & -1.905 \\
         & 0.540  & 0.849 & 0.091 & -3.768  & -1.799 & -1.855 \\
         & 0.544  & 0.834 & 0.103 & -3.666  & -1.751 & -1.806 \\
         & 0.549  & 0.814 & 0.114 & -3.566  & -1.703 & -1.759 \\
         & 0.553  & 0.791 & 0.129 & -3.469  & -1.656 & -1.712 \\
         & 0.558  & 0.770 & 0.148 & -3.373  & -1.611 & -1.666 \\
         & 0.562  & 0.732 & 0.174 & -3.280  & -1.566 & -1.622 \\
critical point     & 0.575  & 0.444 & 0.444  & -3.025  & -1.444 & -1.500 \\   
\hline

  1.20\footnotemark[6] 
         & 0.562  & 0.903 & 0.029 & -5.414  & -2.585 & -2.589 \\
         & 0.572  & 0.890 & 0.035 & -5.129  & -2.449 & -2.453 \\
         & 0.583  & 0.874 & 0.042 & -4.859  & -2.320 & -2.324 \\
         & 0.594  & 0.858 & 0.051 & -4.601  & -2.197 & -2.201 \\
         & 0.604  & 0.837 & 0.062 & -4.355  & -2.079 & -2.083 \\
         & 0.616  & 0.813 & 0.077 & -4.121  & -1.968 & -1.972 \\
         & 0.627  & 0.784 & 0.096 & -3.898  & -1.861 & -1.865 \\
         & 0.638  & 0.748 & 0.121 & -3.685  & -1.760 & -1.764 \\
         & 0.650   & 0.692 & 0.162 & -3.482  & -1.663 & -1.667 \\
critical point     & 0.672  & 0.418 & 0.418  & -3.133  & -1.496 & -1.500 \\   
\hline

  1.25\footnotemark[6] 
         & 0.557  & 0.910 & 0.008 & -7.940  & -3.791 & -3.728 \\
         & 0.572  & 0.900 & 0.011 & -7.371  & -3.519 & -3.456 \\
         & 0.588  & 0.888 & 0.014 & -6.842  & -3.267 & -3.204 \\
         & 0.604  & 0.877 & 0.019 & -6.352  & -3.033 & -2.969 \\
         & 0.621  & 0.862 & 0.024 & -5.896  & -2.815 & -2.752 \\
         & 0.638  & 0.846 & 0.032 & -5.471  & -2.612 & -2.549 \\
         & 0.656  & 0.820 & 0.041 & -5.075  & -2.423 & -2.360 \\
         & 0.674  & 0.797 & 0.054 & -4.705  & -2.247 & -2.183 \\
         & 0.693  & 0.766 & 0.073 & -4.360  & -2.082 & -2.018 \\
         & 0.712  & 0.730 & 0.098 & -4.037  & -1.928 & -1.864 \\
         & 0.732  & 0.671 & 0.142 & -3.735  & -1.783 & -1.720 \\ 
critical point     & 0.766  & 0.395 & 0.395 & -3.374  & -1.563 & -1.500 \\   
\hline

  1.30\footnotemark[6] 
         & 0.579  & 0.882 & 0.005 & -9.483  & -4.528 & -4.436 \\
         & 0.597  & 0.874 & 0.007 & -8.772  & -4.188 & -4.097 \\
         & 0.616  & 0.864 & 0.009 & -8.117  & -3.875 & -3.784 \\
         & 0.635  & 0.850 & 0.012 & -7.512  & -3.587 & -3.495 \\
         & 0.654  & 0.838 & 0.015 & -6.953  & -3.320 & -3.228 \\
         & 0.675  & 0.822 & 0.021 & -6.435  & -3.073 & -2.981 \\
         & 0.696  & 0.805 & 0.027 & -5.955  & -2.843 & -2.752 \\
         & 0.717  & 0.787 & 0.035 & -5.509  & -2.631 & -2.539 \\
         & 0.739  & 0.761 & 0.047 & -5.095  & -2.433 & -2.342 \\
         & 0.762  & 0.735 & 0.063 & -4.710  & -2.249 & -2.157 \\
         & 0.786  & 0.700 & 0.084 & -4.350  & -2.077 & -1.986 \\
         & 0.810  & 0.649 & 0.124 & -4.015  & -1.917 & -1.826 \\
critical point     & 0.868  & 0.370  & 0.370  & -3.333  & -1.591 & -1.500 \\   
\hline
 
\end{tabular}
\footnotetext[6]{This work}
\end{table}
\endgroup

\begingroup
\squeezetable
\begin{table}
\caption{Coexistence properties and second virial coeficients for square-well 
fluids with different $\lambda$ (continued from TABLE \ref{table5}). See the 
article for symbol definitions. }

\label{table6}

\begin{tabular}{ccccccc}
\hline\hline
\hspace{0.5cm}$\lambda$ \hspace{0.5cm} &\hspace{0.5cm}$T^{*}$\hspace{0.5cm} & 
\hspace{0.5cm}$\rho_{L}^{*}$\hspace{0.5cm} &
\hspace{0.5cm}$\rho_{V}^{*}$\hspace{0.5cm} & \hspace{0.5cm}$B_2$\hspace{0.5cm} 
& 
\hspace{0.5cm}$B_2^{*}$\hspace{0.5cm} & \hspace{0.5cm}$B_{2s}$\hspace{0.5cm}  \\

\hline

  1.50\footnotemark[7] & 0.900  & 0.693 & 0.013 & -8.042  & -3.840 & -3.317 \\
         & 0.950  & 0.671 & 0.020 & -7.183  & -3.430 & -2.907 \\
         & 1.000  & 0.647 & 0.030 & -6.453  & -3.081 & -2.558 \\
         & 1.050  & 0.620 & 0.044 & -5.824  & -2.781 & -2.258 \\
         & 1.100  & 0.586 & 0.063 & -5.278  & -2.520 & -1.997 \\
         & 1.120  & 0.570 & 0.073 & -5.079  & -2.425 & -1.902 \\
critical point     & 1.218 & 0.302 & 0.302 & -4.237  & -2.023 & -1.500   \\   
\hline

  1.75\footnotemark[8] & 1.400  & 0.595 & 0.024 & -7.426  & -3.546 & -2.826 \\
         & 1.550  & 0.532 & 0.045 & -6.180  & -2.951 & -2.231 \\
         & 1.650  & 0.493 & 0.079 & -5.513  & -2.632 & -1.912 \\
         & 1.730  & 0.435 & 0.109 & -5.050  & -2.411 & -1.691 \\
critical point     & 1.808 & 0.265 & 0.265 & -4.649  & -2.220 & -1.500   \\ 
\hline

  2.00\footnotemark[7] & 2.100  & 0.642 & 0.024 & -6.848 & -3.270    & -2.681 \\
         & 2.150  & 0.619 & 0.028 & -6.588 & -3.145  & -2.557 \\
         & 2.250  & 0.577 & 0.038 & -6.110 & -2.917  & -2.329 \\
         & 2.350  & 0.533 & 0.054 & -5.682 & -2.713  & -2.124 \\
         & 2.450  & 0.487 & 0.072 & -5.295 & -2.528  & -1.940 \\
critical point     & 2.736 & 0.235 & 0.235 & -4.374 & -2.089  & -1.500   \\  
\hline

  2.30\footnotemark[9] & 2.60  & 0.847 & 0.002 & -8.876  & -4.238 & -3.910 \\
         & 2.70  & 0.829 & 0.004 & -8.390  & -4.006 & -3.678 \\
         & 2.80 & 0.811 & 0.005 & -7.945  & -3.793 & -3.465 \\
         & 2.90  & 0.794 & 0.008 & -7.536  & -3.598 & -3.270 \\
         & 3.00  & 0.775 & 0.009 & -7.158  & -3.418 & -3.090 \\
         & 3.10  & 0.756 & 0.013 & -6.809  & -3.251 & -2.923 \\
         & 3.20  & 0.734 & 0.016 & -6.485  & -3.096 & -2.769 \\
         & 3.30  & 0.712 & 0.020 & -6.184  & -2.953 & -2.625 \\
         & 3.50  & 0.664 & 0.033 & -5.640  & -2.693 & -2.365 \\
         & 3.60  & 0.633 & 0.042 & -5.394  & -2.576 & -2.248 \\
         & 3.70  & 0.603 & 0.052 & -5.163  & -2.465 & -2.137 \\
         & 3.80  & 0.574 & 0.063 & -4.946  & -2.362 & -2.034 \\
         & 3.90  & 0.566 & 0.075 & -4.742  & -2.264 & -1.936 \\
         & 3.95  & 0.542 & 0.080 & -4.644  & -2.217 & -1.889 \\
         & 4.00  & 0.516 & 0.085 & -4.548  & -2.172 & -1.844 \\
         & 4.10  & 0.482 & 0.112 & -4.366  & -2.085 & -1.757 \\
         & 4.15  & 0.476 & 0.137 & -4.278  & -2.043 & -1.715 \\
         & 4.20  & 0.447 & 0.146 & -4.193  & -2.002 & -1.674 \\
critical point     & 4.43  & 0.266 & 0.266 & -3.828  & -1.828 & -1.500 \\   
\hline
  
  2.50\footnotemark[9] & 3.30  & 0.783 & 0.003 & -8.747  & -4.177 & -3.945 \\
         & 3.40  & 0.772 & 0.003 & -8.379  & -4.001 & -3.769 \\
         & 3.50  & 0.760 & 0.005 & -8.035  & -3.837 & -3.605 \\
         & 3.60  & 0.749 & 0.006 & -7.713  & -3.683 & -3.451 \\
         & 3.70  & 0.735 & 0.008 & -7.411  & -3.538 & -3.307 \\
         & 3.90  & 0.709 & 0.012 & -6.858  & -3.275 & -3.043 \\
         & 4.00  & 0.697 & 0.013 & -6.605  & -3.154 & -2.922 \\
         & 4.10  & 0.682 & 0.016 & -6.366  & -3.040 & -3.808 \\
         & 4.20  & 0.669 & 0.020 & -6.140  & -2.932 & -3.700 \\
         & 4.30  & 0.654 & 0.024 & -5.925  & -2.829 & -2.598 \\
         & 4.40  & 0.636 & 0.028 & -5.722  & -2.732 & -2.500 \\
         & 4.60  & 0.607 & 0.037 & -5.344  & -2.551 & -2.320 \\
         & 4.80  & 0.570 & 0.051 & -5.000  & -2.387 & -2.156 \\
         & 5.00  & 0.529 & 0.069 & -4.687  & -2.238 & -2.007 \\
         & 5.10  & 0.530 & 0.077 & -4.541  & -2.168 & -1.937 \\
         & 5.20  & 0.512 & 0.090 & -4.401  & -2.101 & -1.870 \\
         & 5.30  & 0.487 & 0.105 & -4.266  & -2.037 & -1.805 \\
         & 5.35  & 0.466 & 0.108 & -4.201  & -2.006 & -1.774 \\
         & 5.40   & 0.459 & 0.125 & -4.137  & -1.975 & -1.744 \\
         & 5.50   & 0.390 & 0.141 & -4.013  & -1.916 & -1.685 \\
critical point   & 5.84  & 0.248 & 0.248 & -3.626  & -1.731 & -1.500 \\   \hline
 
\end{tabular}
\footnotetext[7]{Data taken from J. Chem. Phys. {\bf 120}, 11754 (2004)}
\footnotetext[8]{Data taken from Mol. Phys. {\bf 100}, 2531 (2002)}
\footnotetext[9]{Data taken from Mol. Phys. {\bf 103}, 129 (2005)}
\end{table}
\endgroup

\begingroup
\squeezetable
\begin{table}
\caption{Coexistence properties and second virial coeficients for square-well 
fluids with different $\lambda$ (continued from TABLE \ref{table6}). See the 
article for symbol definitions. }

\label{table7}

\begin{tabular}{ccccccc}
\hline\hline
\hspace{0.5cm}$\lambda$ \hspace{0.5cm} &\hspace{0.5cm}$T^{*}$\hspace{0.5cm} & 
\hspace{0.5cm}$\rho_{L}^{*}$\hspace{0.5cm} &
\hspace{0.5cm}$\rho_{V}^{*}$\hspace{0.5cm} & \hspace{0.5cm}$B_2$\hspace{0.5cm} 
& 
\hspace{0.5cm}$B_2^{*}$\hspace{0.5cm} & \hspace{0.5cm}$B_{2s}$\hspace{0.5cm}  \\

\hline

  2.70\footnotemark[9] & 4.00  & 0.778 & 0.002 & -9.019  & -4.306 & -4.055 \\
         & 4.10  & 0.767 & 0.003 & -8.714  & -4.161 & -3.910 \\
         & 4.20  & 0.758 & 0.003 & -8.425  & -4.023 & -3.772 \\
         & 4.30  & 0.743 & 0.004 & -8.151  & -3.892 & -3.641 \\
         & 4.40  & 0.736 & 0.005 & -7.890  & -3.767 & -3.516 \\
         & 4.50  & 0.725 & 0.007 & -7.643  & -3.649 & -3.398 \\
         & 4.60  & 0.713 & 0.008 & -7.407  & -3.537 & -3.286 \\
         & 4.70  & 0.704 & 0.009 & -7.183  & -3.430 & -3.179 \\
         & 4.80  & 0.694 & 0.010 & -6.969  & -3.327 & -3.076 \\
         & 4.90  & 0.680 & 0.012 & -6.764  & -3.230 & -2.979 \\
         & 5.00  & 0.672 & 0.014 & -6.569  & -3.136 & -2.886 \\
         & 5.20  & 0.651 & 0.018 & -6.203  & -2.962 & -2.711 \\
         & 5.40  & 0.628 & 0.025 & -5.866  & -2.801 & -2.550 \\
         & 5.50  & 0.616 & 0.028 & -5.708  & -2.725 & -2.474 \\
         & 5.60  & 0.606 & 0.032 & -5.556  & -2.653 & -2.402 \\
         & 6.40  & 0.506 & 0.071 & -4.523  & -2.160 & -1.909 \\
         & 6.50  & 0.494 & 0.081 & -4.413  & -2.107 & -1.856 \\
         & 6.60  & 0.486 & 0.096 & -4.307  & -2.056 & -1.806 \\
         & 6.70  & 0.454 & 0.100 & -4.204  & -2.007 & -1.756 \\
         & 6.80  & 0.440 & 0.119 & -4.105  & -2.960 & -1.709 \\
         & 6.90  & 0.422 & 0.134 & -4.008  & -2.914 & -1.663 \\
critical point  & 7.28  & 0.242 & 0.242 & -3.667  & -1.751 & -1.500  \\ \hline

  3.00\footnotemark[9] & 6.50  & 0.722 & 0.010 & -6.962  & -3.324 & -3.093 \\   
         & 6.60  & 0.710 & 0.011 & -6.814  & -3.253 & -3.022 \\
         & 6.70  & 0.700 & 0.012 & -6.671  & -3.185 & -2.954 \\
         & 6.80  & 0.692 & 0.014 & -6.532  & -3.119 & -2.887 \\
         & 6.90  & 0.680 & 0.016 & -6.398  & -3.055 & -2.823 \\
         & 7.00  & 0.669 & 0.017 & -6.268  & -2.993 & -2.761 \\
         & 7.10  & 0.657 & 0.019 & -6.142  & -2.932 & -2.701 \\
         & 7.20  & 0.646 & 0.020 & -6.019  & -2.874 & -2.642 \\
         & 7.30  & 0.639 & 0.023 & -5.900  & -2.817 & -2.586 \\
         & 7.40  & 0.627 & 0.026 & -5.785  & -2.762 & -2.530 \\
         & 8.60  & 0.523 & 0.065 & -4.620  & -2.206 & -1.974 \\
         & 8.80  & 0.502 & 0.077 & -4.459  & -2.129 & -1.897 \\
         & 9.00  & 0.477 & 0.089 & -4.305  & -2.055 & -1.824 \\
         & 9.20  & 0.443 & 0.100 & -4.158  & -1.985 & -1.754 \\
         & 9.40  & 0.418 & 0.122 & -4.018  & -1.918 & -1.687 \\
         & 9.45  & 0.408 & 0.125 & -3.984  & -1.902 & -1.671 \\
critical point  & 10.01  & 0.248 & 0.248 & -3.627  & -1.732 & -1.500  \\ \hline
 
\end{tabular}
\footnotetext[9]{Data taken from Mol. Phys. {\bf 103}, 129 (2005)}
\end{table}
\endgroup

\end{document}